\documentclass[10pt]{amsart}
\usepackage {amssymb}
\usepackage {amsmath}

\newcommand{\Real} {\mathbb R}
\newcommand{\Complex} {\mathbb C}
\newcommand{\Set} {\mathbb S}
\newcommand{\Operator} {\mathbb P}
\newcommand{\DiffOperator} {\mathbb L}

\begin{document}

\title[Differential Transfer Matrix Method]
{Analytical solution of linear ordinary differential equations by
differential transfer matrix method}

\author{Sina Khorasani and Ali Adibi}

\address{School of Electrical and Computer Engineering,
Georgia Institute of Technology, Atlanta, GA 30332-0250, Fax:
(404) 894-4641, Email: sina.khorasani@ece.gatech.edu}

\date{March 11, 2003}

\begin{abstract}

We report a new analytical method for exact solution of
homogeneous linear ordinary differential equations with arbitrary
order and variable coefficients. The method is based on the
definition of jump transfer matrices and their extension into
limiting differential form. The approach reduces the $n$th-order
differential equation to a system of $n$ linear differential
equations with unity order. The full analytical solution is then
found by the perturbation technique. The important feature of the
presented method is that it deals with the evolution of
independent solutions, rather than its derivatives. We prove the
validity of method by direct substitution of the solution in the
original differential equation. We discuss the general properties
of differential transfer matrices and present several analytical
examples, showing the applicability of the method. We show that
the Abel-Liouville-Ostogradski theorem can be easily recovered
through this approach.

\end{abstract}


\maketitle

\section{Introduction}

Solution of ordinary differential equations (ODEs) is in general possible by
different methods [1]. Lie group analysis [2-6] can be effectively used in
cases when a symmetry algebra exists. Factorization methods are reported for
reduction of ODEs into linear autonomous forms [7,8] with constant
coefficients, which can be readily solved. But these methods require some
symbolic transformations and are complicated for general higher order ODEs,
and thus not suitable for many computational purposes where the form of ODE
is not available in a simple closed form.

General second-order linear ODEs can be solved in terms of special
function [9] and analytic continuation [10]. Spectral methods
[11,12] and orthogonal polynomial [13] solutions of linear ODEs
have been also reported. Solution properties of linear ODEs with
arbitrary orders are discussed in a report [14], and several
approximation methods have been also published [15]. All of these
methods express the solution in terms of infinite series, relying
mainly on numerical computations. Analytic solution of such
equations has been also recently presented [16], but only at a
singular point.

Linear ODEs of the $n$th-order can also be transformed to a system
of $n$ linear first-order differential equations by reduction
method [17], where the derivative of one parameter is regarded as
the next parameter. This approach is based on the evolution of the
solution function and its derivates up to the order $n$. This
method is therefore not useful when the evolution of independent
solutions is of interest rather than the derivatives, such as
waves in inhomogeneous media.

Recently, we have developed a new matrix-based exact analytical
method in order to study the propagation of electromagnetic waves
in non-uniform media with an arbitrary refractive index function
[18,19]. The method is based on the modification of the well-known
transfer matrix theory in optics [20] and quantum mechanics [21],
and its extension into differential form. This method is simple,
exact, and efficient, while reflecting the basic properties of the
physical system described by the differential equation. We have
also developed proper formalism for anisotropic non-uniform media,
in which the governing equations are no longer scalar [22]. In
this method, the solution is found by integration of an exponent
square matrix and then taking its matrix exponential.

In this paper, we show that the differential transfer matrix
method (DTMM) [18,19] can be also used for exact solution of
linear ODEs with arbitrary variable coefficients. We only consider
the homogeneous equation and its linear independent solutions,
since once the linear independent solutions are known, the
particular solution can be found by the method of variation of
parameters due to Lagrange [2]. We show that the $n$th order ODE
is transformed into a system of $n$ homogeneous first order linear
equations, which can be integrated either numerically by
Runge-Kutta [23,24] or reflexive [25] methods, or analytically by
matrix exponential techniques. Through direct substitution we
rigorously show that the presented analytical solution satisfies
the corresponding differential equation, and justify the approach
through several illustrative examples. We also show that the
famous Abel-Liouville-Ostogradsky theorem for the Wronskian of a
linear ODE [26-28] can be easily proved through the DTMM.

In $\S 2$ we describe the outline of the problem and then
formulate the jump transfer matrix method in $\S 3$. In $\S 4$
differential formulation, the fundamental theorem of DTMM, and
treatment of singularities are described. Finally, the application
of the DTMM to several ODEs of various orders is presented.

\section{Statement of Problem}

Consider the homogeneous ordinary linear differential equation of order
$n$

\begin{equation}
\label{eq1} \DiffOperator f\left( x \right) = 0, \quad f:\Set
\mapsto \Complex
\end{equation}

\noindent where $f\left( x \right)$ is an unknown analytic complex
function in the set of complex variables $\Complex$ with the
connected domain $\Set \subset \Complex$, and $\DiffOperator$ is a
linear operator given by

\begin{equation}
\label{eq2} \DiffOperator = \sum\limits_{m = 0}^n {a_m \left( x
\right)\frac{d^m}{dx^m}} , \quad a_n \left( x \right) \equiv 1.
\end{equation}

Here we assume that $a_m :\Set \mapsto \Complex,m = 0,\ldots, n -
1$ are arbitrary analytic complex functions of $x$, being referred
to as variable coefficients. If coefficients are constant, then
under the condition described just below, a general solution to
(\ref{eq1}) is [1,29]

\begin{equation}
\label{eq3} f\left( x \right) = c_1 \exp \left( {k_1 x} \right) +
c_2 \exp \left( {k_2 x} \right) + \cdots + c_n \exp \left( {k_n x}
\right),
\end{equation}

\noindent where $c_i ,i = 1,\ldots, n$ are complex constants
determined from either boundary or initial conditions, and $k_i ,i
= 1,\ldots, n$, being referred here to as wavenumbers, satisfy the
characteristic equation

\begin{equation}
\label{eq4} \sum\limits_{m = 0}^n {a_m k_i ^m} = 0,\quad i =
1,\ldots, n.
\end{equation}

Moreover, for (\ref{eq3}) to be a general solution of (\ref{eq1})
with constant coefficients, it is necessary that $\forall i,j;k_i
\ne k_j$, i.e. (\ref{eq4}) has exactly $n$ distinct roots. Now, if
coefficient functions are not constant, then we propose a general
solution to (\ref{eq1}) of the form

\begin{equation}
\label{eq5} f\left( x \right) = f_1 \left( x \right)\exp \left[
{k_1 \left( x \right) x} \right] + f_2 \left( x \right) \exp
\left[ {k_2 \left( x \right) x} \right] + \cdots + f_n \left( x
\right)\exp \left[ {k_n \left( x \right)x} \right],
\end{equation}

\noindent
or equivalently

\begin{equation}
\label{eq6} f\left( x \right) = \exp \left[ {{\rm {\bf \Phi
}}\left( x \right)} \right]^{\,t}\,{\rm {\bf F}}\left( x \right),
\end{equation}

\begin{equation}
\label{eq7} {\rm {\bf F}}\left( x \right) = \left[
{{\begin{array}{*{20}c}
 {f_1 \left( x \right)} \hfill \\
 {f_2 \left( x \right)} \hfill \\
 \vdots \hfill \\
 {f_n \left( x \right)} \hfill \\
\end{array} }} \right],\\
{\rm {\bf \Phi }}\left( x \right) = \left[ {{\begin{array}{*{20}c}
 {k_1 \left( x \right)x} \hfill \\
 {k_2 \left( x \right)x} \hfill \\
 \vdots \hfill \\
 {k_n \left( x \right)x} \hfill \\
\end{array} }} \right],
\end{equation}

\noindent where the superscript $t$ denotes the transposed matrix.
Hereinafter, we shall refer to ${\rm {\bf F}}\left( x \right)$ and
${\rm {\bf \Phi }}\left( x \right)$ respectively as envelope and
phase vector functions. In the above equation $f_i \left( x
\right)$ are functions to be determined and $k_i(x)$, being
referred to as wavenumber functions, are complex functions of $x$
satisfying the generalized algebraic characteristic equation

\begin{equation}
\label{eq8} \sum\limits_{m = 0}^n {a_m \left( x \right)\,k_i
^{\,m}\left( x \right)} = 0,\quad i = 1,\ldots, n.
\end{equation}

For the sake of convenience, we here define the exponential of a
vector ${\rm {\bf v}} = \left\{ {v_i } \right\}_{n\times 1}$ as
$\exp \left( {\rm {\bf v}} \right) = \left\{ {\exp \left( {v_i }
\right)} \right\}_{n\times 1}$. In contrast, $\exp \left( {\rm
{\bf M}} \right)$ when \textbf{M} is a square matrix represents
the matrix exponential of \textbf{M}, given by

\begin{equation}
\label{eq8a} \exp \left( {\rm {\bf M}} \right) = {\rm {\bf I}} +
\sum\limits_{m = 1}^\infty {\frac{1}{m!}{\rm {\bf M}}^m}.
\end{equation}

We define a domain $\Set$ to be non-degenerate if (\ref{eq8}) has
exactly $n$ distinct roots for all $x \in \Set$, and $m$-fold
degenerate if it is non-degenerate everywhere, but at finite
number of isolated points, being referred to as singularities, at
which (\ref{eq8}) has a minimum of $n-m$ distinct roots. If $\Set$
is at least 1-fold degenerate for all $x \in \Set$, then $\Set$ is
referred to as entirely degenerate. Later, we shall observe that
non-singular solutions require $\Set$ to be non-degenerate, and
degenerate domains require special treatment.

We also define the diagonal wavenumber matrix as

\begin{equation}
\label{eq9} {\rm {\bf K}}\left( x \right) = \left[ {k_i \left( x
\right)\delta _{ij} } \right]_{\,n\times n} = {\rm diag} \left[
k_1(x), \ldots, k_n(x) \right],
\end{equation}

\noindent with $\delta _{ij}$ being the Kronecker delta.
Obviously, the diagonal elements of ${{\rm {\bf K}}\left( x
\right)}$, and thus ${{\rm {\bf K}}\left( x \right)}$ itself must
satisfy (\ref{eq8}). Moreover, $\left| {{\rm {\bf K}}\left( x
\right)} \right| = a_0 \left( x \right)$ and ${\rm tr}\left\{
{{\rm {\bf K}}\left( x \right)} \right\} = a_{n - 1} \left( x
\right)$, where ${\rm tr}\left\{\cdot\right\}$ denotes the trace
operator. The phase vector and wavenumber matrix are thus related
as ${\rm {\bf \Phi }}\left( x \right) = x{\rm {\bf K}}\left( x
\right)\,{\rm {\bf 1}}_{n\times 1}$, where ${\bf 1}_{n\times 1}$
is a vector with unity elements. Hence, one has $\exp \left[ {{\rm
{\bf \Phi }}\left( x \right)} \right] = \exp \left[ {x{\rm {\bf
K}}\left( x \right)} \right]\,{\rm {\bf 1}}_{n\times 1}$.

\section{Jump Transfer Matrices}

\subsection{Transfer Matrix of a Single Jump}

Without loss of generality, suppose that $\Set \subset \Real$ with
$\Real$ being the set of real numbers. Furthermore, we let the
variable coefficients $a_i(x),i=1,\ldots, n$ to be such stepwise
constant discontinuous functions at $x=X$ that ${\rm {\bf
K}}\left( {x < X} \right) = {\rm {\bf K}}_A$ and ${\rm {\bf
K}}\left( {x > X} \right) = {\rm {\bf K}}_B$ become constant
matrices. We furthermore accept that both the sub-spaces $\Set_A =
\left\{ {\left. x \right|x \in \Set,x < X} \right\}$ and $\Set_B =
\left\{ {\left. x \right|x \in \Set,x > X} \right\}$ are
non-degenerate. In this case, the solution in $\Set$ would be
given by

\begin{equation}
\label{eq10} f\left( x \right) = \left\{ {{\begin{array}{*{20}c}
 {\exp \left[ {{\rm {\bf \Phi }}_A \left( x \right)} \right]^{\,t}\,{\rm
{\bf F}}_A ,\quad x \in \Set_A } \hfill \\
 {\exp \left[ {{\rm {\bf \Phi }}_B \left( x \right)} \right]^{\,t}\,{\rm
{\bf F}}_B ,\quad x \in \Set_B } \hfill \\
\end{array} }} \right.,
\end{equation}

\noindent where the envelope vectors ${\rm {\bf F}}_A$ and ${\rm
{\bf F}}_B$ must be constant, and ${\rm {\bf \Phi }}_A \left( x
\right) = \left\{ {_A k_i x} \right\}_{n\times 1}$, ${\rm {\bf
\Phi }}_B \left( x \right) = \left\{ {_B k_i x} \right\}_{n\times
1}$ with ${ }_Ak_i$ and ${ }_Bk_i$ satisfying the characteristic
equation (\ref{eq4}) in $\Set_A$ and $\Set_B$. Analyticity of
$f\left( x \right)$ across $x = X$ requires that

\begin{equation}
\label{eq11} f^{\left( m \right)}\left( {X^ - } \right) =
f^{\left( m \right)}\left( {X^ + } \right),\quad m = 0,\ldots, n -
1,
\end{equation}

\noindent where $f^{\left( m \right)}\left( x \right) $ is the
$m$th derivative of $f\left( x \right)$ with respect to $x$.
Expanding (\ref{eq11}) results in the set of equations

\begin{equation}
\label{eq12} {\rm {\bf D}}_B \exp \left( {X{\rm {\bf K}}_B }
\right){\rm {\bf F}}_B = {\rm {\bf D}}_A \exp \left( {X{\rm {\bf
K}}_A } \right){\rm {\bf F}}_A,
\end{equation}

\noindent in which ${\rm {\bf K}}_A$ and ${\rm {\bf K}}_B$ are
given in (\ref{eq9}). Also ${\rm {\bf D}}_A = \left[ {{ }_Ak_j ^{i
- 1}} \right]_{\,n\times n}$ and ${\rm {\bf D}}_B = \left[ {{
}_Bk_j ^{i - 1}} \right]_{\,n\times n}$ are matrices of
Vandermonde type [30] given by

\begin{equation}
\label{eq13} {\rm {\bf D}}_A = \left[ {{\begin{array}{*{20}c}
 1 \hfill & 1 \hfill & \cdots \hfill & 1 \hfill \\
 {_A k_1 } \hfill & {_A k_2 } \hfill & \cdots \hfill & {_A k_n } \hfill \\
 \vdots \hfill & \vdots \hfill & \hfill & \vdots \hfill \\
 {_A k_1 ^{n - 1}} \hfill & {_A k_2 ^{n - 1}} \hfill & \cdots \hfill & {_A
k_n ^{n - 1}} \hfill \\
\end{array} }} \right],
{\rm {\bf D}}_B = \left[ {{\begin{array}{*{20}c}
 1 \hfill & 1 \hfill & \cdots \hfill & 1 \hfill \\
 {_B k_1 } \hfill & {_B k_2 } \hfill & \cdots \hfill & {_B k_n } \hfill \\
 \vdots \hfill & \vdots \hfill & \hfill & \vdots \hfill \\
 {_B k_1 ^{n - 1}} \hfill & {_B k_2 ^{n - 1}} \hfill & \cdots \hfill & {_B
k_n ^{n - 1}} \hfill \\
\end{array} }} \right].
\end{equation}

One can rewrite (\ref{eq12}) as

\begin{equation}
\label{eq14} {\rm {\bf F}}_B = {\rm {\bf Q}}_{A \to B} {\rm {\bf
F}}_A,
\end{equation}

\begin{equation}
\label{eq15} {\rm {\bf Q}}_{A \to B} = \exp \left( { - X{\rm {\bf
K}}_B } \right)\,{\rm {\bf D}}_B ^{ - 1}{\rm {\bf D}}_A \exp
\left( {X{\rm {\bf K}}_A } \right),
\end{equation}

\noindent in which ${\rm {\bf Q}}_{A \to B}$ is defined as the
jump transfer matrix from $A$ to $B$. Similarly, one can define
the downward jump transfer matrix as ${\rm {\bf Q}}_{B \to A} =
{\rm {\bf Q}}_{A \to B}^{ - 1}$.

The determinant of the jump transfer matrix ${\rm {\bf Q}}_{A \to
B}$ is

\begin{equation}
\label{eq16} \left| {{\rm {\bf Q}}_{A \to B} } \right| = \exp
\left[ {X\left( {{\rm tr}\left\{ {{\rm {\bf K}}_A } \right\} -
{\rm tr}\left\{ {{\rm {\bf K}}_B } \right\}} \right)}
\right]\frac{\left| {{\rm {\bf D}}_A } \right|}{\left| {{\rm {\bf
D}}_B } \right|},
\end{equation}

\noindent in which we have used the identity $\left|\exp({\bf
M})\right|=\exp({\rm tr}\left\{ {\bf M}\right\} )$. The
determinants can be also evaluated by noting the fact that $\left|
{{\rm {\bf D}}_r } \right|$ are determinants of Vandermonde
matrices, given by [30]

\begin{equation}
\label{eq17} \left| {{\rm {\bf D}}_r } \right| = \left\{
{{\begin{array}{*{20}c}
 {\begin{array}{l}
 1,\quad n = 1 \\
 \\
 \end{array}} \hfill \\
 {\prod\limits_{i > j} {\left( {{ }_rk_i - { }_rk_j } \right)} ,\quad n > 1}
\hfill \\
\end{array} }} \right.\; r = A,B.
\end{equation}

\noindent Here, the product operator runs over all possible
ordered pairs $(i,j)$ with $i>j$. Now since ${\rm tr}\left\{ {{\rm
{\bf K}}_r } \right\} = \sum\limits_{i = 1}^n {{ }_rk_i } = {
}_ra_{n - 1}$ with $r = A,B$, (\ref{eq16}) can be expanded as

\begin{equation}
\label{eq18} \left| {{\rm {\bf Q}}_{A \to B} } \right| = \exp
\left[ X ({ }_Aa_{n - 1}-{}_Ba_{n - 1})\right]\prod\limits_{i > j}
{\frac{\left( {{ }_Ak_i - { }_Ak_j } \right)}{\left( {{ }_Bk_i - {
}_Bk_j } \right)}}.
\end{equation}

\subsection{Transfer Matrix of Multiple Jumps}

When variable coefficients $a_i(x),i=1,\ldots ,n$ are stepwise
functions of $x$ with arbitrary number of discontinuities or jumps
over interfaces, ${\rm {\bf K}}\left( x \right)$ can be expressed
as

\begin{equation}
\label{eq19} {\rm {\bf K}}\left( x \right) = {\rm {\bf K}}_p
,\quad x \in \Set_p ,\quad p = 0, \ldots, P ,
\end{equation}

\noindent where the subset $\Set_p = \left\{ {\left. x \right|X_p
< x < X_{p + 1} } \right\},p = 0,\ldots, P$ is referred to as the
$p$th layer. Obviously, $X_0 = \inf \left\{\Set \right\} = \inf
\left\{ {\Set_0 } \right\}$ and $X_{P + 1} = \sup \left\{\Set
\right\} = \sup \left\{ {\Set_P } \right\}$. In this case, the
corresponding envelope vectors are related as

\begin{equation}
\label{eq20} {\rm {\bf F}}_s = {\rm {\bf Q}}_{r \to s} {\rm {\bf
F}}_r ,\quad r,s = 0, \ldots, P,
\end{equation}

\noindent where ${\rm {\bf Q}}_{r \to s}$ is the transfer matrix
from layer $r$ to $s$, obtained by multiplication of single jump
transfer matrices as

\begin{equation}
\label{eq21} {\rm {\bf Q}}_{r \to s} = {\rm {\bf Q}}_{s - 1 \to s}
{\rm {\bf Q}}_{s - 2 \to s - 1} \cdots {\rm {\bf Q}}_{r \to r +
1}.
\end{equation}

\subsection{Properties of Jump Transfer Matrix }

The transfer matrix ${\rm {\bf Q}}_{r \to s}$ satisfies the basic
properties

\begin{equation}
\label{eq21a} {\rm {\bf Q}}_{r \to r} = {\rm {\bf
I}},\;(self-projection)
\end{equation}

\begin{equation}
\label{eq21b}{\rm {\bf Q}}_{r \to s} = {\rm {\bf Q}}_{s \to r}^{ -
1},\;(inversion)
\end{equation}

\begin{equation}
\label{eq21c} {\rm {\bf Q}}_{r \to s} = {\rm {\bf Q}}_{u \to s}
{\rm {\bf Q}}_{r \to u},\; (decomposition)
\end{equation}

\begin{equation}
\label{eq21d} \left| {{\rm {\bf Q}}_{r \to s} } \right| = \exp
\left[ {\sum\limits_{p = r}^s {X_p \sum\limits_{i = 1}^n {\left(
{{ }_pk{ }_i - { }_{p + 1}k{ }_i} \right)} } }
\right]\prod\limits_{i > j} {\frac{\left( {{ }_rk_i - { }_rk_j }
\right)}{\left( {{ }_sk_i - { }_sk_j } \right)}},\;(determinant)
\end{equation}

\noindent with $r,s,u = 0, \ldots, P$. The properties
(\ref{eq21a}), (\ref{eq21b}), and (\ref{eq21c}) are direct results
of the definition of transfer matrices, and the determinant
property (\ref{eq21d}) follows (\ref{eq21}) and (\ref{eq18}).

From (\ref{eq18}) it can be observed that for systems with
vanishing $_r a_{n - 1} ,r = 0, \ldots, P$, the determinant
property simplifies into

\begin{equation}
\label{eq22} \left| {{\rm {\bf Q}}_{r \to s} } \right| =
\prod\limits_{i > j} {\frac{\left( {{ }_rk_j - { }_rk_i }
\right)}{\left( {{ }_sk_j - { }_sk_i } \right)}}.
\end{equation}

\noindent This expression has the important feature of being
dependent only on the local properties of the initial layer $r$
and target layer $s$.

The transfer matrix ${\rm {\bf Q}}_{r \to s}$ satisfies the
scaling property, that is it remains unchanged under
transformations $X_p \to \alpha X_p$ and ${ }_rk_i \to \alpha ^{ -
1}{ }_rk_i$. This can be directly observed either from (\ref{eq1})
by making the change of variable $x \to \alpha x$, which
accordingly scales the roots in an inverse manner, or direct
substitution in (\ref{eq15}). It also preserves the shifting
property

\begin{equation}
\label{eq22a} {\rm {\bf Q}}_{r \to s} = \exp \left( { - \xi {\rm
{\bf K}}_s } \right)\;{\rm {\bf \hat {Q}}}_{r \to s} \exp \left(
{\xi {\rm {\bf K}}_r } \right),\;(shifting)
\end{equation}

\noindent in which $\xi$ is the amount of shift over $x$-axis and
${\rm {\bf \hat {Q}}}_{r \to s}$ is the transfer matrix
corresponding to shifted space. This property follows (\ref{eq15})
and (\ref{eq21}). The shift theorem plays an important role in the
theory of wave propagation in one-dimensional periodic structures
in electromagnetic and optics [31].

In the next section we shall show that how one can solve (\ref{eq1}) in its most
general form by extension of jump transfer matrices into differential form.

\section{Differential Transfer Matrix Method (DTMM)}

\subsection{Formulation }

We first assume that the wavenumber matrix ${\rm {\bf K}}\left( x
\right)$ is a smoothly varying function of $x$. For the moment, we
also assume that $\Set$ is non-degenerate. Following the notation
in the previous section, let $A$ and $B$ denote respectively the
sub-domains $\Set_A = \left[ {x - \delta x,x} \right[$ and $\Set_B
= \left[ {x,x + \delta x} \right[$. If $\delta x$ is small enough,
one can approximate the wavenumber matrix as ${\rm {\bf K}}\left(
{t \in \Set_A } \right) \approx {\rm {\bf K}}\left( x \right)
\equiv {\rm {\bf K}}_A$, and ${\rm {\bf K}}\left( {t \in \Set_B }
\right) \approx {\rm {\bf K}}\left( {x + \delta x} \right) \equiv
{\rm {\bf K}}_B$. Accordingly, one has

\begin{equation}
\label{eq23} {\rm {\bf F}}\left( {x + \delta x} \right) \approx
{\rm {\bf F}}_B = {\rm {\bf Q}}_{A \to B} {\rm {\bf F}}_A \approx
{\rm {\bf Q}}_{A \to B} {\rm {\bf F}}\left( x \right)
\end{equation}

The above equation is accurate to the first order. So, one can
expand the transfer matrix ${\rm {\bf Q}}_{A \to B}$ through its
definition (\ref{eq15}) as

\begin{equation}
\label{eq24} \begin{array}{l}
 {\rm {\bf Q}}_{A \to B} = \exp \left( { - x{\rm {\bf K}}_B } \right)\,{\rm
{\bf D}}_B ^{ - 1}{\rm {\bf D}}_A \exp \left( {x{\rm {\bf K}}_A } \right) \\
 \quad \quad \;\; = \exp \left[ { - x\left( {{\rm {\bf K}}_A + \delta {\rm
{\bf K}}} \right)} \right]\,( {{\rm {\bf D}}_A + \delta {\rm {\bf
D}}}
)^{ - 1}{\rm {\bf D}}_A \exp \left( {x{\rm {\bf K}}_A } \right) \\
 \quad \quad \;\; \approx \exp \left( { - x{\rm {\bf K}}_A } \right)\left(
{{\rm {\bf I}} - x\delta {\rm {\bf K}}} \right)\,\left[ {{\rm {\bf
D}}_A ^{ - 1} - {\rm {\bf D}}_A ^{ - 1}(\delta {\rm {\bf D}}){\rm
{\bf D}}_A ^{ - 1}} \right]{\rm
{\bf D}}_A \exp \left( {x{\rm {\bf K}}_A } \right) \\
 \quad \quad \;\; = \exp \left( { - x{\rm {\bf K}}_A } \right)\left( {{\rm
{\bf I}} - x\delta {\rm {\bf K}}} \right)\,\left( {{\rm {\bf I}} -
{\rm {\bf D}}_A ^{ - 1}\delta {\rm {\bf D}}} \right)\exp \left(
{x{\rm {\bf K}}_A }
\right) \\
 \quad \quad \;\; \approx \exp \left( { - x{\rm {\bf K}}_A } \right)\left(
{{\rm {\bf I}} - x\delta {\rm {\bf K}} - {\rm {\bf D}}_A ^{ -
1}\delta {\rm
{\bf D}}} \right)\exp \left( {x{\rm {\bf K}}_A } \right) \\
 \end{array},
\end{equation}

\noindent in which we have used the Taylor expansion of matrix
exponential and differential property of inverse matrices [32],
and have neglected the 2nd- and higher-order variations. The
matrix $\delta {\rm {\bf K}}$ can be approximated as

\begin{equation}
\label{eq25} \delta {\rm {\bf K}} = \left[ {\delta k_i \delta
_{ij} } \right] \approx \left[ {{k}'_i \left( x \right)\delta
_{ij} } \right] \delta x = \frac{d}{dx}{\rm {\bf K}}\left( x
\right)\delta x.
\end{equation}

\noindent From (\ref{eq13}) one also has

\begin{equation}
\label{eq26} \begin{array}{l}
 \delta {\rm {\bf D}} = \left[ {{\begin{array}{*{20}c}
 0 \hfill & 0 \hfill & \cdots \hfill & 0 \hfill \\
 1 \hfill & 1 \hfill & \cdots \hfill & 1 \hfill \\
 \vdots \hfill & \vdots \hfill & \hfill & \vdots \hfill \\
 {\left( {n - 1} \right)_A k_1 ^{n - 2}} \hfill & {\left( {n - 1} \right)_A
k_2 ^{n - 2}} \hfill & \cdots \hfill & {\left( {n - 1} \right)_A
k_n ^{n -
2}} \hfill \\
\end{array} }} \right]\frac{d}{dx}{\rm {\bf K}}\left( x \right)\delta x \\
 \quad \equiv {\rm {\bf C}}\delta {\rm {\bf K}} \\
 \end{array}
.
\end{equation}

\noindent So we can rewrite (\ref{eq23}) as

\begin{equation}
\label{eq27} \delta {\rm {\bf F}}\left( x \right) \approx \left(
{{\rm {\bf Q}}_{A \to B} - {\rm {\bf I}}} \right){\rm {\bf
F}}\left( x \right) \equiv {\rm {\bf U}}\left( x \right){\rm {\bf
F}}\left( x \right)\delta x.
\end{equation}

\noindent But the matrix ${\rm {\bf U}}\left( x \right)$ can be
found from (\ref{eq24}) and (\ref{eq26}) as

\begin{equation}
\label{eq28} {\rm {\bf U}}\left( x \right) \approx - \exp \left( {
- x{\rm {\bf K}}} \right)\left( {x{\rm {\bf I}} + {\rm {\bf D}}^{
- 1}{\rm {\bf C}}} \right)\frac{\delta {\rm {\bf K}}}{\delta x
}\exp \left( {x{\rm {\bf K}}} \right) ,
\end{equation}

\noindent where the trivial subscript $A$ has been dropped.

Now, if we let $\delta x$ approach zero, (\ref{eq27}) and
(\ref{eq28}) become exact and can be rewritten as

\begin{equation}
\label{eq29} d{\rm {\bf F}}\left( x \right) = {\rm {\bf U}}\left(
x \right){\rm {\bf F}}\left( x \right)dx,
\end{equation}
\begin{equation}
\label{eq30} {\rm {\bf U}}\left( x \right) = - x{\rm {\bf
{K}'}}\left( x \right) - \exp \left[ { - x{\rm {\bf K}}\left( x
\right)} \right]{\rm {\bf D}}\left( x \right)^{ - 1}{\rm {\bf
C}}\left( x \right){\rm {\bf {K}'}}\left( x \right)\exp \left[
{x{\rm {\bf K}}\left( x \right)} \right].
\end{equation}

\noindent Hereinafter, we refer to ${\rm {\bf U}}\left( x \right)$
as the differential transfer or the kernel matrix. In
(\ref{eq30}), we have

\begin{equation}
\label{eq32} {\rm {\bf C}}\left( x \right) = \left[ {\left( {i -
1} \right)k_j^{i - 2} \left( x \right)} \right]_{n\times n}, {\rm
{\bf D}}\left( x \right) = \left[ {k_j^{i - 1} \left( x \right)}
\right]_{n\times n},
\end{equation}
\begin{equation}
{\rm {\bf K}}\left( x \right) = \left[ {k_i \left( x \right)\delta
_{ij} } \right]_{n\times n}, {\rm {\bf {K}'}}\left( x \right) =
\left[ {{k}'_i \left( x \right)\delta _{ij} } \right]_{n\times n}.
\end{equation}

While (\ref{eq29}) can be integrated by numerical methods [23-25],
it permits an exact solution of the form

\begin{equation}
\label{eq310} {\rm {\bf F}}\left( {x_2 } \right) = {\rm {\bf
Q}}_{x_1 \to x_2 } {\rm {\bf F}}\left( {x_1 } \right),
\end{equation}

\noindent for $\forall x_1 ,x_2 \in \Set.$. Here ${\rm {\bf
Q}}_{x_1 \to x_2 }$ is referred to as the transfer matrix from
$x_1$ to  $x_2$, given by the Dyson's perturbation theory as
[33-35]

\begin{equation}
\label{eq311} \begin{array}{l}

{\rm {\bf Q}}_{x_1 \to x_2 } ={\bf I} + \int\limits_{x_1
}^{x_2}dt_1{\rm {\bf U}}(t_1) + \int\limits_{x_1 }^{x_2}dt_1
\int\limits_{x_1 }^{t_1}dt_2{\rm {\bf U}}(t_1){\rm {\bf U}}(t_2) +
\ldots + \\ \quad \quad \int\limits_{x_1 }^{x_2}dt_1
\int\limits_{x_1 }^{t_1}dt_2 \int\limits_{x_1 }^{t_2}dt_3 \ldots
\int\limits_{x_1 }^{t_{m-1}}dt_m {\rm {\bf U}}(t_1) \ldots {\rm
{\bf U}}(t_m) + \ldots,
\end{array}
\end{equation}

\noindent By defining $\Operator$ as the so-called chronological
ordering operator [33-35], it is possible to rewrite (\ref{eq311})
as
\begin{equation}
\label{eq311a} {\rm {\bf Q}}_{x_1 \to x_2 } =
\sum_{m=0}^{\infty}\frac{1}{m!} \int\limits_{x_1 }^{x_2}dt_1
\int\limits_{x_1 }^{x_2}dt_2 \int\limits_{x_1 }^{x_2}dt_3 \ldots
\int\limits_{x_1 }^{x_2}dt_m \Operator \left[ {\rm {\bf U}}(t_1)
\ldots {\rm {\bf U}}(t_m) \right].
\end{equation}

\noindent Often, (\ref{eq311a}) is symbolically written as [33-35]

\begin{equation}
\label{eq31} {\rm {\bf Q}}_{x_1 \to x_2 }  =  \Operator \exp
\left[ {\int\limits_{x_1 }^{x_2 } {{\rm {\bf U}}\left( x
\right)dx} } \right] \equiv \Operator \exp \left( {{\rm {\bf
M}}_{x_1 \to x_2 } } \right),
\end{equation}

\noindent in which ${\rm {\bf M}}_{x_1 \to x_2}$ is the integral
of the kernel matrix and referred to as the transfer exponent
matrix. The above expression can be greatly simplified if the
kernel matrix ${\rm {\bf U}}(x)$ satisfies one of the few existing
sufficient conditions for integrability, including when it
commutes with ${\rm {\bf M}}_{x_1 \to x_2}$ [36-38], or it
satisfies the Fedorov's condition [38], or it commutes with itself
for $\forall x_1, x_2 \in \Set$. In this case, one has

\begin{equation}
\label{eq313} {\rm {\bf Q}}_{x_1 \to x_2 }  = \exp \left[
{\int\limits_{x_1 }^{x_2 } {{\rm {\bf U}}\left( x \right)dx} }
\right].
\end{equation}

Although in general (\ref{eq313}) is not necessarily the exact
transfer matrix, it is known that both (\ref{eq310}) and
(\ref{eq313}) must have the same determinant [39]. Also their
traces are equal at least to the second order. This can be easily
justified by comparing the expansion of (\ref{eq313}) to
(\ref{eq311}) and using the cyclic permutation property of ${\rm
tr}\left\{\cdot\right\}$ [30]. This means for the case of
second-order equations with $n=2$, (\ref{eq313}) can be used
instead of (\ref{eq31}), with an accuracy better than
second-order. If needed, the exact transfer matrix can be directly
found by numerical integration of equation [39]

\begin{equation}
\label{eq314} d{\rm {\bf Q}}_{x_1 \to x }  = {\rm {\bf U}}\left( x
\right) {\rm {\bf Q}}_{x_1 \to x } dx,
\end{equation}

\noindent with the initial condition ${\rm {\bf Q}}_{x_1 \to x_1
}={\bf I}$.

\subsection{Properties of Transfer Matrix}

The transfer matrix ${\rm {\bf Q}}_{x_1 \to x_2 }$ from $x_1$ to
$x_2$ clearly preserves the properties (\ref{eq21a}),
(\ref{eq21b}) and (\ref{eq21c}), and as discussed above, its
determinant is always equal to the determinant of its counterpart
given by (\ref{eq313}). Therefore, $\left|{\rm {\bf Q}}_{x_1 \to
x_2}\right|=\exp({\rm tr}\left\{{\rm {\bf M}}_{x_1 \to x_2 }
\right\})$. But the transfer exponent matrix ${\rm {\bf M}}_{x_1
\to x_2 }$ can be written as the sum of the jump ${\rm {\bf
J}}_{x_1 \to x_2 }$ and propagation ${\rm {\bf T}}_{x_1 \to x_2 }$
matrices given by

\begin{equation}
\label{eq33} {\rm {\bf J}}_{x_1 \to x_2 } = - \int\limits_{x_1
}^{x_2 } {\exp \left[ { - x{\rm {\bf K}}\left( x \right)}
\right]{\rm {\bf D}}\left( x \right)^{ - 1}\frac{d}{dx}\left\{
{{\rm {\bf D}}\left( x \right)\exp \left[ {x{\rm {\bf K}}\left( x
\right)} \right]} \right\}\,dx},
\end{equation}

\begin{equation}
\label{eq34} {\rm {\bf T}}_{x_1 \to x_2 } = \int\limits_{x_1
}^{x_2 } {{\rm {\bf K}}\left( x \right)dx}.
\end{equation}

\noindent The validity of the identity ${\rm {\bf M}}_{x_1 \to x_2
}={\rm {\bf J}}_{x_1 \to x_2 }+{\rm {\bf T}}_{x_1 \to x_2 }$ can
be verified by comparing the integrands of both sides. The
propagation matrix ${\rm {\bf T}}_{x_1 \to x_2 }$ is diagonal and
also has a diagonal matrix exponential given by

\begin{equation}
\label{eq35} \exp \left( {{\rm {\bf T}}_{x_1 \to x_2 } } \right) =
\left[ {\exp \left( {\int\limits_{x_1 }^{x_2 } {k_i \left( x
\right)dx} } \right)\delta _{ij} } \right].
\end{equation}

\noindent Therefore, its determinant can be easily found by
multiplying its diagonal elements. The determinant $\left|{\rm
{\bf J}}_{x_1 \to x_2 }\right|$ can be obtained using the relation
${\rm {\bf {D}'}}\left( x \right) = {\rm {\bf C}}\left( x
\right){\rm {\bf {K}'}}\left( x \right)$ according to
(\ref{eq26}), and the identity (see Appendix A)

\begin{equation}
\label{eq36} \left| \exp \left[ {\int\limits_{x_1}^{x_2} {{\rm
{\bf H}}\left( x \right)^{ - 1}{\rm {\bf {H}'}}\left( x \right)dx}
} \right] \right| = \left| {\rm {\bf H}}^{ - 1}\left( x_1
\right){\rm {\bf H}}\left( x_2 \right) \right|,
\end{equation}

\noindent
as

\begin{equation}
\label{eq37} \left|\exp \left( {{\rm {\bf J}}_{x_1 \to x_2 } }
\right) \right|= \left| \exp \left[ { - x_2 {\rm {\bf K}}\left(
{x_2 } \right)} \right]{\rm {\bf D}}\left( {x_2 } \right)^{ -
1}{\rm {\bf D}}\left( {x_1 } \right)\exp \left[ {x_1 {\rm {\bf
K}}\left( {x_1 } \right)} \right] \right|.
\end{equation}

Finally, the determinant $\left| {\rm {\bf Q}}_{x_1 \to x_2
}\right|$ can be thus found using the well-known identities [29]
$\left| {\exp \left( {{\rm {\bf A}} + {\rm {\bf B}}} \right)}
\right| = \left| {\exp \left( {\rm {\bf A}} \right)} \right|\times
\left| {\exp \left( {\rm {\bf B}} \right)} \right|$, and $\left|
{\exp \left( {\rm {\bf M}} \right)} \right| = \exp \left( {{\rm
tr}\left\{ {\rm {\bf M}} \right\}} \right)$ as

\begin{equation}
\label{eq38} \left| {{\rm {\bf Q}}_{x_1 \to x_2 } } \right| = \exp
\left( {{\rm tr}\left\{ {{\rm {\bf T}}_{x_1 \to x_2 } } \right\}}
\right)\times \exp \left( {{\rm tr}\left\{ {{\rm {\bf J}}_{x_1 \to
x_2 } } \right\}} \right).
\end{equation}

\noindent Using the (\ref{eq35}), (\ref{eq37}), and (\ref{eq17})
one can finally obtain

\begin{equation}
\label{eq39}
\begin{array}{l}
 \left| {{\rm {\bf Q}}_{x_1 \to x_2 } } \right| \\
  \quad \quad = \exp \left( {x_1 {\rm tr}\left\{
{{\rm {\bf K}}\left( {x_1 } \right)} \right\} - x_2 {\rm
tr}\left\{ {{\rm {\bf K}}\left( {x_2 } \right)} \right\} -
\int\limits_{x_1 }^{x_2 } {{\rm tr}\left\{ {{\rm {\bf K}}\left( x
\right)} \right\}dx} } \right)\times \prod\limits_{i
> j} {\frac{k_i \left( {x_1 } \right) - k_j \left( {x_1 } \right)}{k_i
\left( {x_2 } \right) - k_j \left( {x_2 } \right)}} \\
 \quad \quad = \exp \left[ {x_1 a_{n - 1} \left( {x_1 } \right) - x_2
a_{n - 1} \left( {x_2 } \right) - \int\limits_{x_1 }^{x_2 } {a_{n
- 1} \left( x \right)dx} } \right]\times \prod\limits_{i > j}
{\frac{k_i \left( {x_1 } \right) - k_j \left( {x_1 } \right)}{k_i
\left( {x_2 } \right) - k_j
\left( {x_2 } \right)}} \\
 \end{array}.
\end{equation}

\noindent This equation can be compared to (\ref{eq21d}) as the
determinant property of transfer matrix. Notice that if $a_{n - 1}
\left( x \right) \equiv 0$, then (\ref{eq39}) reduces to

\begin{equation}
\label{eq40} \left| {{\rm {\bf Q}}_{x_1 \to x_2 } } \right| =
\prod\limits_{i > j} {\frac{k_i \left( {x_1 } \right) - k_j \left(
{x_1 } \right)}{k_i \left( {x_2 } \right) - k_j \left( {x_2 }
\right)}}.
\end{equation}

\noindent This shows that the determinant of the transfer matrix
is only a function of starting and end points $x_1$ and $x_2$.
Note that the transformation

\begin{equation}
\label{eq41} f\left( x \right) \to \exp \left[ { - \textstyle{1
\over n}\int {a_{n - 1} \left( x \right)dx} } \right] h\left( x
\right)
\end{equation}

\noindent in (\ref{eq1}) always gives rise to a new $n$th-order
ODE with identically vanishing coefficient of $h^{\left( {n - 1}
\right)}\left( x \right)$ term. Therefore, the above property of
determinants can be always met through the abovementioned
transformation.

The shifting property (23) and scaling properties must also
preserve, since the transfer matrix ${\rm {\bf Q}}_{x_1 \to x_2
}$is indeed obtained by dividing the sub-domain $\left[ {x_1 ,x_2
} \right]$ into infinitely many thin layers and multiplying the
infinitesimal jump transfer matrices.

\subsection{Justification of DTMM }

Here we prove that the DTMM formulation is exact by showing that
the solution satisfies (\ref{eq1}) through direct substitution.

\textbf{Lemma 1. }\textit{Let the Vandermonde matrix
}\textbf{D}\textit{ be invertible. Then the elements of its
inverse }${\rm {\bf D}}^{ - 1} = \left[ {\gamma _{ij} }
\right]$\textit{ satisfy }$\sum\limits_{i = 1}^n {\gamma _{ir} k_i
^{m - 1}} = \delta _{mr}$\textit{ where }$1 \le m \le n$.

\textbf{Proof. }The Vandermonde coefficients $\gamma _{ij}$ are
found from the expansion of Lagrange interpolation polynomials as
[40,41]

\begin{equation}
\label{eq42} \Gamma _i \left( t \right) = \prod\limits_{j \ne i}
{\frac{t - k_j }{k_i - k_j }} = \sum\limits_{j = 1}^n {\gamma
_{ij} t^{j - 1}} ,\quad i = 1,\ldots, n.
\end{equation}

\noindent Obviously $\Gamma _i \left( {k_j } \right) =
\sum\limits_{r = 1}^n {\gamma _{ir} k_j ^{r - 1}} = \delta _{ij}$.
Therefore $k_j ^{m - 1} = \sum\limits_{i = 1}^n {k_i ^{m -
1}\Gamma _i \left( {k_j } \right)}$, or

\begin{equation}
\label{eq43} k_j ^{m - 1} = \sum\limits_{i = 1}^n {k_i ^{m -
1}\sum\limits_{r = 1}^n {\gamma _{ir} k_j ^{r - 1}} } =
\sum\limits_{r = 1}^n {\left( {\sum\limits_{i = 1}^n {\gamma _{ir}
k_i ^{m - 1}} } \right)k_j ^{r - 1}} \equiv \sum\limits_{r = 1}^n
{a_{mr} k_j ^{r - 1}}.
\end{equation}

\noindent Given a fixed $m$, the above equation can be transformed
into a linear system of equations as ${\rm {\bf A}}^t{\rm {\bf D}}
= {\rm {\bf B}}^t$, with \textbf{D} being the Vandermonde matrix,
${\rm {\bf A}} = \left[ {a_{rm} } \right]$ being the column vector
of unkowns, and ${\rm {\bf B}} = \left[ {k_i ^{m - 1}} \right]$
being the input column vector. Since \textbf{D} is invertible by
assumption, then \textbf{A} and thus $a_{rm}$ must be unique.
However, (\ref{eq43}) is satisfied through setting $a_{rm} =
\delta _{rm}$ for $1 \le m \le n$, and hence the proof is
complete. {\raggedright $\Box$}

\textbf{Lemma 2 (Derivative Lemma).} \textit{Suppose that }${\rm
{\bf F}}\left( x \right)$\textit{ is an }$n\times 1$\textit{
vector function, defined on the connected non-degenerate domain
}$\Set$\textit{, satisfying the differential equation
}(\ref{eq29})\textit{ with the kernel matrix given by
}(\ref{eq30})\textit{. Then, we have}

\begin{equation}
\label{eq44} \frac{d^m}{dx^m}\left\{ {\exp \left[ {{\rm {\bf \Phi
}}\left( x \right)} \right]^{\,t}{\rm {\bf F}}\left( x \right)}
\right\} = \exp \left[ {{\rm {\bf \Phi }}\left( x \right)}
\right]^{\,t}{\rm {\bf K}}^m\left( x \right){\rm {\bf F}}\left( x
\right), m = 0,\ldots, n,
\end{equation}

\noindent \textit{where }${\rm {\bf K}}\left( x \right)$\textit{
is the wavenumber matrix given by }(\ref{eq9})\textit{ and }${\rm
{\bf \Phi }}\left( x \right)$\textit{ is the corresponding phase
vector defined in} (\ref{eq7}) \textit{.}

\textbf{Proof.} For $m = 0$ the proof is trivial. Also for $m \ge
0$ we have

\begin{equation}
\label{eq45}
\begin{array}{l}
 \frac{d}{dx}\left\{ {\exp \left( {\rm {\bf \Phi }} \right)^{\,t}{\rm {\bf
K}}^m{\rm {\bf F}}} \right\} \\
 \quad = \frac{d}{dx}\exp \left( {\rm {\bf \Phi }} \right)^{\,t}{\rm {\bf
K}}^m{\rm {\bf F}} + \exp \left( {\rm {\bf \Phi }}
\right)^{\,t}\frac{d}{dx}{\rm {\bf K}}^m{\rm {\bf F}} + \exp
\left( {\rm
{\bf \Phi }} \right)^{\,t}{\rm {\bf K}}^m\frac{d}{dx}{\rm {\bf F}} \\
 \quad = \exp \left( {\rm {\bf \Phi }} \right)^{\,t}\left( {{\rm {\bf K}}^{m
+ 1} + x{\rm {\bf K}}^m{\rm {\bf {K}'}} + m{\rm {\bf K}}^{m -
1}{\rm {\bf
{K}'}} + {\rm {\bf K}}^m{\rm {\bf U}}} \right){\rm {\bf F}} \\
 \end{array},
\end{equation}

\noindent where the dependence of matrices on $x$ is not shown for
the sake of convenience. From the definition of the kernel matrix
(\ref{eq30}), (\ref{eq45}) can be simplified as

\begin{equation}
\label{eq46}
\begin{array}{l}
 \frac{d}{dx}\left\{ {\,\exp \left( {\rm {\bf \Phi }} \right)^{\,t}{\rm {\bf
K}}^m{\rm {\bf F}}} \right\} \\
 \quad \quad = \exp \left( {\rm {\bf \Phi }} \right)^{\,t}{\rm {\bf
 K}}^{m+1}{\rm {\bf F}} +
\\ \quad \quad \quad
\exp \left( {\rm {\bf \Phi }} \right)^{\,t}\left[
{m{\rm {\bf K}}^{m - 1}{\rm {\bf {K}'}} - {\rm {\bf K}}^m\exp
\left( { - x{\rm {\bf K}}} \right){\rm {\bf D}}^{ - 1}{\rm {\bf
C{K}'}}\exp \left( {x{\rm {\bf
K}}} \right)} \right]\,{\rm {\bf F}} \\
 \quad \quad = \exp \left( {\rm {\bf \Phi }} \right)^{\,t}{\rm {\bf K}}^{m +
1}{\rm {\bf F}} + {\rm {\bf 1}}_{n\times 1}^{\,t} \left( {m{\rm
{\bf K}}^{m - 1} - {\rm {\bf K}}^m{\rm {\bf D}}^{ - 1}{\rm {\bf
C}}} \right)\,{\rm {\bf
{K}'}}\exp \left( {x{\rm {\bf K}}} \right){\rm {\bf F}} \\
 \quad \quad \equiv \exp \left( {\rm {\bf \Phi }} \right)^{\,t}{\rm {\bf
K}}^{m + 1}{\rm {\bf F}} + {\rm {\bf 1}}_{1\times n} {\rm {\bf
V{K}'}}\exp
\left( {x{\rm {\bf K}}} \right){\rm {\bf F}} \\
 \end{array}
,
\end{equation}

\noindent where ${\rm {\bf 1}}_{n\times 1}$ is defined under
(\ref{eq9}) and ${\rm {\bf V} }\equiv {m{\rm {\bf K}}^{m - 1} -
{\rm {\bf K}}^m{\rm {\bf D}}^{ - 1}{\rm {\bf C}}}$ . Now we show
that the row vector ${\rm {\bf 1}}_{1\times n} {\rm {\bf V}}$ is
identically zero, but only if $m < n$. By defining the elements of
the inverse of the Vandermonde matrix as ${\rm {\bf D}}^{ - 1} =
\left[ {\gamma _{ij} } \right]$, we have

\begin{equation}
\label{eq47}
\begin{array}{l}
 {\rm {\bf 1}}_{1\times n} {\rm {\bf V}} = m{\rm {\bf 1}}_{1\times n} {\rm
{\bf K}}^{m - 1} - {\rm {\bf 1}}_{1\times n} {\rm {\bf K}}^m{\rm
{\bf D}}^{ - 1}{\rm {\bf C}} \\ \quad \quad = \left[
{\sum\limits_{i = 1}^n {mk_i ^{m - 1}\delta _{ij} } } \right] -
\left[ {\sum\limits_{i = 1}^n {\sum\limits_{r = 1}^n {\gamma
_{ir} \left( {r - 1} \right)k_j ^{r - 2}k_i ^m} } } \right] \\
 \quad \quad \, = \left[ {mk_j ^{m - 1} - \sum\limits_{r = 1}^{n - 1} {rk_j
^{r - 1}\sum\limits_{i = 1}^n {\gamma _{ir + 1} k_i ^m} } } \right] \\
 \end{array}.
\end{equation}

\noindent Since the inverse of \textbf{D} must exist by
assumption, we thus have according to Lemma 1

\begin{equation}
\label{eq48} {\rm {\bf 1}}_{1\times n} {\rm {\bf V}} = \left[
{mk_j ^{m - 1} - \sum\limits_{r = 1}^{n - 1} {rk_j ^{r - 1}\delta
_{mr} } } \right] = {\rm {\bf 0}}_{1\times n},\; m < n .
\end{equation}

\noindent Therefore, the proof of theorem follows immediately by
induction. {\raggedright $\Box$}

\textbf{Theorem 1 (Fundamental theorem of differential transfer
matrix method).} \textit{Suppose that }${\rm {\bf F}}\left( x
\right)$\textit{ satisfies the derivative Lemma, with the kernel
matrix }${\rm {\bf K}}\left( x \right)$\textit{ satisfying the
characteristic equation }(\ref{eq8})\textit{. Then, the function
}$f\left( x \right)$\textit{ defined by }(\ref{eq6})\textit{ is a
solution of the differential equation }(\ref{eq1})$. $

\textbf{Proof.} By expansion of the operator $\DiffOperator$ given
in (\ref{eq2}) on $f\left( x \right)$ given in (\ref{eq6}), and
using the derivative Lemma we have

\begin{equation}
\label{eq49}
\begin{array}{l}
 {\rm \DiffOperator}f\left( x \right) = \sum\limits_{m = 0}^n {a_m \left( x
\right)\frac{d^m}{dx^m}} f\left( x \right)\\ \quad =
\sum\limits_{m = 0}^n {a_m \left( x \right)\frac{d^m}{dx^m}\left\{
{\exp \left[ {{\rm {\bf \Phi
}}\left( x \right)} \right]^{\,t}{\rm {\bf F}}\left( x \right)} \right\}} \\
 \quad = \sum\limits_{m = 0}^n {a_m \left( x \right)\exp \left[ {{\rm {\bf
\Phi }}\left( x \right)} \right]^{\,t}{\rm {\bf K}}^m\left( x
\right){\rm {\bf F}}\left( x \right)}\\ \quad = \exp \left[ {{\rm
{\bf \Phi }}\left( x \right)} \right]^{\,t}\left[ {\sum\limits_{m
= 0}^n {a_m \left( x \right){\rm {\bf
K}}^m\left( x \right)} } \right]{\rm {\bf F}}\left( x \right) \\
 \end{array}.
\end{equation}

\noindent But the summation within the brackets is the null matrix
because of (\ref{eq8}). Thus the right-hand-side of the above
equation must be identically zero. This completes the proof.
{\raggedright $\Box$}

\subsection{Linear Independent Solutions}

Consider the set of vectors ${\rm {\bf F}}_i \in \Set^n,i =
1,\ldots, n$, forming a basis in $\Set^n$. If ${\bf Q}$ is a
non-singular matrix, then the vectors ${\rm {\bf G}}_i = {\rm {\bf
QF}}_i \in \Complex^n,i = 1,\ldots, n$ would clearly constitute
another basis on $\Set^n$. Let $\Set$ be a non-degenerate domain
with the differential equation (\ref{eq1}) and solution given by
(\ref{eq310}). Since $\Set$ is non-degenerate, then ${\rm {\bf
Q}}_{x_1 \to x_2 }$ must be non-singular for all $x_1 ,x_2 \in
\Set$. Therefore the set of $n$ vector functions defined by ${\rm
{\bf G}}_i \left( x \right) = {\rm {\bf Q}}_{a \to x} {\rm {\bf
F}}_i \left( a \right) \in \Set^n$ form a basis on $\Set^n$. Let
$\Psi $ denote the set of all functions operating on $\Set$, as
$\Psi = \left\{ {\psi :\Set \mapsto \Complex} \right\}$. We define
the scalar functions $g_i \in \Psi$ as $g_i \left( x \right) =
\exp \left[ {{\rm {\bf \Phi }}\left( x \right)} \right]^{\,t}{\rm
{\bf G}}_i \left( x \right)$ and show that they form a basis, by
inspection of their Wronskian determinant.

\textbf{Theorem 2. }\textit{With the above definitions, the set of
functions }$g_i \left( x \right), i = 1,\ldots, n$\textit{ form n
linear independent solutions on $\Psi $}.

\textbf{Proof. }The Wronskian matrix in $\Psi$ can be written as
${\rm {\bf W}} = \left[ {g_i ^{\left( {j - 1} \right)}\left( x
\right)} \right]$. Using Theorem 1 one has ${\rm {\bf W}} = \left[
{\exp \left( {\rm {\bf \Phi }} \right){\rm {\bf K}}^{j - 1}{\rm
{\bf Q}}_{a \to x} {\rm {\bf F}}_i } \right]$, with the
dependences on $x$ omitted. Now we define the $n\times n$ matrix
${\rm {\bf F}} = \left[ {{\rm {\bf F}}_i } \right]$. Then ${\rm
{\bf W}} = {\rm {\bf D}}\exp \left( {x{\rm {\bf K}}} \right)\,{\rm
{\bf Q}}_{a \to x} \,{\rm {\bf F}}$. Because $\Set$ is
non-degenerate $\left| {\rm {\bf D}} \right| \ne 0$. Also, ${\rm
{\bf Q}}_{a \to x}$ must be non-singular and thus $\left| {\rm
{\bf Q}}_{a \to x}\right| \ne 0$. Now since according to the
assumption, ${\rm {\bf F}}_i$ form a basis on $\Set^n$, $\left|
{\rm {\bf F}} \right| \ne 0$ and therefore for the Wronskian
determinant we have $\left| {\rm {\bf W}} \right| \ne 0$. Thus,
the proof is complete. {\raggedright $\Box$}

\subsection{Treatment of Singularities}

Now let $\Set$ is degenerate at a finite number of isolated
singularities $\xi _i \in \Set,i = 1,\ldots, \Sigma$, at which
$\left| {{\rm {\bf D}}\left( {\xi _i } \right)} \right| = 0$.
Therefore, the ${\rm {\bf U}}\left( {\xi _i } \right)$ is singular
and ${\rm {\bf M}}_{x_1 \to x_2 }$ does not exist. Without loss of
generality, we can assume $\Sigma = 1$. If $\Set$ represents the
integration domain $\left[ {x_1 ,x_2 } \right]$, the total
transfer matrix ${\rm {\bf Q}}_{x_1 \to x_2 }$ by the
decomposition property (\ref{eq21c}) can be written as

\begin{equation}
\label{eq50} {\rm {\bf Q}}_{x_1 \to x_2 } = {\rm {\bf Q}}_{\xi +
\delta x \to x_2 } {\rm {\bf Q}}_{\xi - \delta x \to \xi + \delta
x} {\rm {\bf Q}}_{x_1 \to \xi - \delta x}.
\end{equation}

\noindent The transfer matrices ${\rm {\bf Q}}_{x_1 \to \xi -
\delta x}$ and ${\rm {\bf Q}}_{\xi + \delta x \to x_2 }$ do exist
and can be evaluated directly by integration and exponentiation of
${\rm {\bf U}}\left( x \right)$, as long as $\delta x$ is positive
and finite. The transfer matrix ${\rm {\bf Q}}_{\xi - \delta x \to
\xi + \delta x} $ enclosing the singularity, also can be
approximated by its equivalent jump transfer matrix as

\begin{equation}
\label{eq51}
\begin{array}{l}
 {\rm {\bf Q}}_{\xi - \delta x \to \xi + \delta x} \approx \exp \left[ { -
\left( {\xi + \delta x} \right){\rm {\bf K}}\left( {\xi + \delta
x} \right)}
\right]\,{\rm {\bf D}}^{ - 1}\left( {\xi + \delta x} \right)\times \\
 \quad \quad \quad \quad \quad \,\,{\rm {\bf D}}\left( {\xi - \delta x}
\right)\exp \left[ {\left( {\xi - \delta x} \right){\rm {\bf
K}}\left( {\xi
- \delta x} \right)} \right] \\
 \end{array}.
\end{equation}

\noindent This approach permits evaluation of the total transfer
matrix ${\rm {\bf Q}}_{x_1 \to x_2 }$ by making finite jumps over
singularites.

If $\Set$ is entirely degenerate, then the transformation $f\left(
x \right) \to w\left( x \right)h\left( x \right)$ with $w\left( x
\right)$ being a non-constant function, results in a completely
different characteristic equation (\ref{eq8}). A proper choice of
$w(x)$ leads to a new $\Set$, which can be no longer entirely
degenerate. Then DTMM can be used to solve for $h\left( x
\right)$.

\section{Examples}

\subsection{First-order ODEs}

It is easy to show the consistency of the method for first order
differential equations with variable coefficients, that is

\begin{equation}
\label{eq55} {f}'\left( x \right) + a_0 \left( x \right)f\left( x
\right) = 0,
\end{equation}

\noindent having the exact solution

\begin{equation}
\label{eq56} f\left( {x_2 } \right) = \exp \left(
{\int\limits_{x_2 }^{x_1 } {a_0 \left( x \right)dx} }
\right)f\left( {x_1 } \right).
\end{equation}

\noindent In this case, the only root of (\ref{eq8}) is
$k(x)=-a_0(x)$, all matrices become scalar with $D = 1$ and $C =
0$, and the transfer exponent reduces to

\begin{equation}
\label{eq57} M_{x_1 \to x_2 } = \int\limits_{x_1 }^{x_2 }
{x\frac{d}{dx}\left[ {a_0 \left( x \right)} \right]dx} = x_2 a_0
\left( {x_2 } \right) - x_1 a_0 \left( {x_1 } \right) -
\int\limits_{x_1 }^{x_2 } {a_0 \left( x \right)dx}.
\end{equation}

\noindent Clearly (\ref{eq313}) holds and therefore, accordingly
one has

\begin{equation}
\label{eq58} F\left( {x_2 } \right) = \exp \left[ {x_2 a_0 \left(
{x_2 } \right) - x_1 a_0 \left( {x_1 } \right) - \int\limits_{x_1
}^{x_2 } {a_0 \left( x \right)dx} } \right]F\left( {x_1 } \right).
\end{equation}

\noindent From (\ref{eq6}), we have

\begin{equation}
\label{eq59} f\left( x \right) = \exp \left[ {xk\left( x \right)}
\right] F\left( x \right).
\end{equation}

\noindent Therefore, using (\ref{eq58}) and (\ref{eq59}) one can
obtain (\ref{eq56}).

\subsection{Second-order ODEs}

In this section, we first consider the simplest second-order
differential equation with variable coefficients, given by

\begin{equation}
\label{eq60} {f}''\left( x \right) + a_0 \left( x \right)f\left( x
\right) = 0.
\end{equation}

\noindent This equation has been studied since 1836 by Sturm [42]
and has been considered in many reports [43-48]. Actually, any
second-order ODE can be rewritten in the form of (\ref{eq60}) by
the transformation given in (\ref{eq41}). Also, the first order
non-linear Riccati equation [2] takes the above form after
suitable transformations.

In physics this equation models the propagation of
transverse-electric (TE) polarized light in one-dimensional (1D)
non-homogeneous media (with a position-dependent refractive
index), or motion of a single electron in a 1D potential trap. In
optical problems $f\left( x \right)$ is the amplitude of
transverse electric field and $a_0 \left( x \right) = k_0
^2\epsilon _r \left( x \right) - \beta ^2$, in which $k_0$ is the
wavenumber of the radiation in vacuum, $\epsilon _r \left( x
\right)$ is the relative permittivity function of the medium, and
$\beta$ is the propagation eigenvalue. In quantum mechanics
$f\left( x \right)$ is the probability wave function and $a_0
\left( x \right) = {2m\left[ {E - V\left( x \right)} \right]}
\mathord{\left/ {\vphantom {{2m\left[ {E - V\left( x \right)}
\right]} {\hbar ^2}}} \right. \kern-\nulldelimiterspace} {\hbar
^2}$, where $m$ is the electron mass, $V\left( x \right)$ is the
electric potential, $\hbar$ is the reduced Planck constant, and
$E$ is the energy. Radial functions of axial gravity waves of a
black hole are governed by a generalized Klein-Gordon or
Regge-Wheeler equation [49,50], which takes a similar form to
(\ref{eq60}).

Here, we consider the solution of this equation by DTMM. For $n =
2$, the kernel matrix ${\rm {\bf U}}\left( x \right)$ can be
simplified from (\ref{eq30}) to

\begin{equation}
\label{eq61} {\rm {\bf U}}\left( x \right) = \left[
{{\begin{array}{*{20}c}
 { - \left( {x + \frac{1}{k_1 - k_2 }} \right){k}'_1 } \hfill &
{\frac{{k}'_2 }{k_2 - k_1 }\exp \left[ { - x\left( {k_1 - k_2 }
\right)}
\right]} \hfill \\
 {\frac{{k}'_1 }{k_1 - k_2 }\exp \left[ { + x\left( {k_1 - k_2 } \right)}
\right]} \hfill & { - \left( {x + \frac{1}{k_2 - k_1 }}
\right){k}'_2 }
\hfill \\
\end{array} }} \right].
\end{equation}

\noindent From (\ref{eq8}), one has $k_i^2(x)+a_0(x)=0,i=1,2$,
that is $k_1 \left( x \right) = - jk\left( x \right)$ and $k_2
\left( x \right) = + jk\left( x \right)$, with $k\left( x \right)
\equiv \sqrt {a_0 \left( x \right)}$. Therefore (\ref{eq61})
simplifies as

\begin{equation}
\label{eq62} {\rm {\bf U}}\left( x \right) = \frac{{k}'\left( x
\right)}{2k\left( x \right)}\left[ {{\begin{array}{*{20}c}
 { - 1 + j2k\left( x \right)x} \hfill & {\exp \left[ {j2xk\left( x \right)}
\right]} \hfill \\
 {\exp \left[ { - 2jxk\left( x \right)} \right]} \hfill & { - 1 - j2k\left(
x \right)x} \hfill \\
\end{array} }} \right],
\end{equation}

\noindent which is in agreement with our previous results
[18,19,22]. The transfer matrix $ {{\rm {\bf Q}}_{x_1 \to x_2 } }$
is then found by (\ref{eq31}) (exact evaluation of matrix
exponential in this case is possible by the theorem described in
the Appendix B). The determinant of $ {{\rm {\bf Q}}_{x_1 \to x_2
} }$ is

\begin{equation}
\label{eq63} \left| {{\rm {\bf Q}}_{x_1 \to x_2 } } \right| =
\frac{k_2 \left( {x_1 } \right) - k_1 \left( {x_1 } \right)}{k_2
\left( {x_2 } \right) - k_1 \left( {x_2 } \right)} = \frac{k\left(
{x_1 } \right)}{k\left( {x_2 } \right)}.
\end{equation}

The complete solution of (\ref{eq60}) is then given by (\ref{eq5})
as

\begin{equation}
\label{eq64} f\left( x \right) = f_1 \left( x \right)\exp \left[ {
- jxk\left( x \right)} \right] + f_2 \left( x \right)\exp \left[ {
+ jxk\left( x \right)} \right].
\end{equation}

The singularities of (\ref{eq60}) correspond to the turning points
at which both wavenumbers become zero. In fact, a singularity
separates the regions in which the waves are evanescent and
propagating, or in other words, $a_0 \left( x \right)$ is
respectively positive or negative. The treatment method for
singularities as described in sub-section 4.5 can be applied to
find the transfer matrix over the singularity, being approximated
by its corresponding jump transfer matrix.

We can classify the singularities depending on the nature of the
waves across the singularity. Here, a singularity at $x = \xi$ is
characterized as $k_1(\xi)=k_2(\xi)$, or $a_0(\xi)=0$.
Correspondingly, the singularity at $x = \xi$ is referred to as
type A if $a_0 \left( {x > \xi } \right)
> 0$ and $a_0 \left( {x < \xi } \right) < 0$, type B if $a_0
\left( {x
> \xi } \right) < 0$ and $a_0 \left( {x < \xi } \right) > 0$, and
otherwise type C. Due to (\ref{eq15}) the jump transfer matrix
from region 1 to 2 across the interface $x = \xi$ for (\ref{eq60})
takes the form [18,19,22,51]

\begin{equation}
\label{eq65} {\rm {\bf Q}}_{1 \to 2} = \left[
{{\begin{array}{*{20}c}
 {\frac{k_2 + k_1 }{2k_2 }e^{ + j\xi \left( {k_2 - k_1 } \right)}} \hfill &
{\frac{k_2 - k_1 }{2k_2 }e^{ + j\xi \left( {k_2 + k_1 } \right)}} \hfill \\
 {\frac{k_2 - k_1 }{2k_2 }e^{ - j\xi \left( {k_2 + k_1 } \right)}} \hfill &
{\frac{k_2 + k_1 }{2k_2 }e^{ - j\xi \left( {k_2 - k_1 } \right)}} \hfill \\
\end{array} }} \right].
\end{equation}

\noindent Here, $k_1^2=-a_0(\xi-\delta x)$ and
$k_1^2=-a_0(\xi+\delta x)$. Therefore, the jump transfer matrix
over singularities given by ${\rm {\bf Q}}_{\xi - \delta x \to \xi
+ \delta x}$ simplifies as

\begin{equation}
\label{eq66a} {\rm {\bf Q}}_{\xi - \delta x \to \xi + \delta x} =
\left[ {{\begin{array}{*{20}c}
 {\frac{1 + j}{2}} \hfill & {\frac{1 - j}{2}} \hfill \\
 {\frac{1 - j}{2}} \hfill & {\frac{1 + j}{2}} \hfill \\
\end{array} }} \right],\;type A
\end{equation}
\begin{equation}
\label{eq66b} {\rm {\bf Q}}_{\xi - \delta x \to \xi + \delta x} =
\left[ {{\begin{array}{*{20}c}
 {\frac{1 - j}{2}} \hfill & {\frac{1 + j}{2}} \hfill \\
 {\frac{1 + j}{2}} \hfill & {\frac{1 - j}{2}} \hfill \\
\end{array} }} \right],\;type B
\end{equation}
\begin{equation}
\label{eq66c}{\rm {\bf Q}}_{\xi - \delta x \to \xi + \delta x} =
\left[ {{\begin{array}{*{20}c}
 1 \hfill & 0 \hfill \\
 0 \hfill & 1 \hfill \\
\end{array} }} \right],\;type C.
\end{equation}

\noindent The results here are obtained by setting $k\left( {\xi
\pm \delta x} \right) \approx \sqrt {\pm \delta x{a}'_0 \left(
{\xi \pm \delta x} \right)}$ through the corresponding Taylor's
expansion of $a_0 \left( x \right)$ around $x = \xi$ and taking
the limit $\delta x \to 0^ +$.

The DTMM has been used to solve (\ref{eq60}) numerically [18,19],
and the results have been compared to other approaches. In cases
that analytical solutions are known, the results have been in
agreement. While the exact transfer matrix is given by
(\ref{eq31}), we have noticed that the reduced form as in
(\ref{eq313}) can be also used and leads to accurate solutions.
Also, energy dispersion of an electron in the 1D potential of
infinite monatomic chain has been computed [22] by this approach.
In general, computation times for DTMM are at least one order of
magnitude lower than the other existing approaches. We also have
discussed how it is possible to extend this approach to
transverse-magnetic (TM) polarized modes and inhomogeneous
anisotropic media, where four field components are solved for at
once [22].

We also have recently exploited the DTMM to periodic
electromagnetic structures [39], in which $a_0(x)$, and thus the
wavenumber functions $k_i(x),i=1,2$ are all periodic. This has led
us to a novel yet simple mathematical description of waves in
inhomogeneous periodic media.

\subsection{Fourth-order ODEs}

As another application example, we obtain the differential
transfer matrix of a simple fourth-order differential equation
that has been discussed in several reports [47,48]

\begin{equation}
\label{eq67} f^{\left( 4 \right)}\left( x \right) + a_0 \left( x
\right)f\left( x \right) = 0.
\end{equation}

\noindent This time from (\ref{eq8}) we have $k_i^4(x)+a_0(x)=0,
i=1,\ldots,4$, and thus $k_1 \left( x \right) = - k\left( x
\right)$, $k_2 \left( x \right) = - jk\left( x \right)$, $k_3
\left( x \right) = + k\left( x \right)$, and $k_4 \left( x \right)
= + jk\left( x \right)$, in which $k\left( x \right) \equiv
\sqrt[4]{ - a_0 \left( x \right)}$. The kernel matrix ${\rm {\bf
U}}\left( x \right)$ is found by {\tt Mathematica} as

\begin{equation}
\label{eq68} \begin{array}{l} {\rm {\bf U}}\left( x \right) =
\frac{{k}'\left( x \right)}{2k\left( x \right)} \times \\
\left[
{{\begin{array}{*{20}c}
 {2xk\left( x \right) - 3} \hfill & {\left( {1 + j} \right)e^{\left( {1 - j}
\right)xk\left( x \right)}} \hfill & {e^{2xk\left( x \right)}}
\hfill & {\left( {1 - j} \right)e^{\left( {1 + j} \right)xk\left(
x \right)}} \hfill
\\
 {\left( {1 - j} \right)e^{\left( {j - 1} \right)xk\left( x \right)}} \hfill
& {2jxk\left( x \right) - 3} \hfill & {\left( {1 + j}
\right)e^{\left( {1 +
j} \right)xk\left( x \right)}} \hfill & {e^{2jxk\left( x \right)}} \hfill \\
 {e^{ - 2xk\left( x \right)}} \hfill & {\left( {1 - j} \right)e^{ - \left(
{1 + j} \right)xk\left( x \right)}} \hfill & { - 2xk\left( x
\right) - 3} \hfill & {\left( {1 + j} \right)e^{\left( {j - 1}
\right)xk\left( x
\right)}} \hfill \\
 {\left( {1 + j} \right)e^{ - \left( {1 + j} \right)xk\left( x \right)}}
\hfill & {e^{ - 2jxk\left( x \right)}} \hfill & {\left( {1 - j}
\right)e^{\left( {1 - j} \right)xk\left( x \right)}} \hfill & { -
2jxk\left(
x \right) - 3} \hfill \\
\end{array} }} \right].
\end{array}
\end{equation}

\noindent Here, it is possible to check the determinant of ${\rm
{\bf Q}}_{x_1 \to x_2 }$ given by

\begin{equation}
\label{eq69} \left| {{\rm {\bf Q}}_{x_1 \to x_2 } } \right| = \exp
\left( {{\rm tr}\left\{ {{\rm {\bf M}}_{x_1 \to x_2 } } \right\}}
\right) = \exp \left( { - 6\int\limits_{x_1 }^{x_2 }
{\frac{{k}'\left( x \right)}{k\left( x \right)}dx} } \right) =
\frac{k^6\left( {x_1 } \right)}{k^6\left( {x_2 } \right)},
\end{equation}

\noindent
in justification of (\ref{eq40}).

To show the applicability of the approach, we take $a{ }_0\left( x
\right) = - a^4x^{ - 4}$ for which (\ref{eq67}) becomes an
Euler-Cauchy equation and therefore has an exact solution given by

\begin{equation}
\label{eq70} f\left( x \right) = c_1 x^{m_1 } + c_2 x^{m_2 } + c_3
x^{m_3 } + c_4 x^{m_4 },
\end{equation}

\noindent where $m_i = \textstyle{3 \over 2}\pm \textstyle{1 \over
2}\sqrt {5\pm 4\sqrt {1 + a^4} } ,i = 1,\ldots, 4$, and $c_i ,i =
1,\ldots, 4$ are arbitrary constants. Since $k\left( x \right) =
ax^{ - 1}$, the kernel matrix (\ref{eq68}) simplifies into

\begin{equation}
\label{eq71} \begin{array}{l}
 {\rm {\bf U}}\left( x \right) \\=  \frac{ - 1}{2x}\left[
{{\begin{array}{*{20}c}
 {2a - 3} \hfill & {\left( {1 + j} \right)e^{\left( {1 - j} \right)a}}
\hfill & {e^{2a}} \hfill & {\left( {1 - j} \right)e^{\left( {1 +
j}
\right)a}} \hfill \\
 {\left( {1 - j} \right)e^{\left( {j - 1} \right)a}} \hfill & {2ja - 3}
\hfill & {\left( {1 + j} \right)e^{\left( {1 + j} \right)a}}
\hfill &
{e^{2ja}} \hfill \\
 {e^{ - 2a}} \hfill & {\left( {1 - j} \right)e^{ - \left( {1 + j} \right)a}}
\hfill & { - 2a - 3} \hfill & {\left( {1 + j} \right)e^{\left( {j
- 1}
\right)a}} \hfill \\
 {\left( {1 + j} \right)e^{ - \left( {1 + j} \right)a}} \hfill & {e^{ -
2ja}} \hfill & {\left( {1 - j} \right)e^{\left( {1 - j} \right)a}}
\hfill &
{ - 2aj - 3} \hfill \\
\end{array} }} \right] \\
 \equiv \frac{3}{2x}{\rm {\bf I}} + \frac{1}{x}{\rm {\bf N}} \\
 \end{array},
\end{equation}

\noindent where \textbf{N} is a traceless constant matrix with
eigenvalues $\lambda _i = \pm \textstyle{1 \over 2}\sqrt {5\pm
4\sqrt {1 + a^4} }$, $i = 1,\ldots, 4$. For this case, the kernel
matrix ${\bf U}(x)$ commutes with the transfer exponent ${\bf
M}_{1\to x}$ so that (\ref{eq313}) becomes exact. By integration
of (\ref{eq71}) we find

\begin{equation}
\label{eq72} {\rm {\bf M}}_{1 \to x} = \frac{3\ln x}{2}\,{\rm {\bf
I}} + \ln x\,{\rm {\bf N}}.
\end{equation}

\noindent Obviously \textbf{I} and \textbf{N} commute, so that

\begin{equation}
\label{eq73} {\rm {\bf Q}}_{1 \to x} = \exp \left( {{\rm {\bf
M}}_{1 \to x} } \right) = x^{3 \mathord{\left/ {\vphantom {3 2}}
\right. \kern-\nulldelimiterspace} 2}\exp \left( {\rm {\bf N}}
\right) = x^{3 \mathord{\left/ {\vphantom {3 2}} \right.
\kern-\nulldelimiterspace} 2}{\rm {\bf R}}\left[ {\exp \left(
{\lambda _i \ln x} \right)\delta _{ij} } \right] {\rm {\bf R}}^{ -
1},
\end{equation}

\noindent where the constant matrix \textbf{R} is the diagonalizer
of \textbf{N}. Therefore, $\left| {{\rm {\bf Q}}_{1 \to x} }
\right| = x^6$, as required by (\ref{eq69}). Furthermore, the
elements of ${\rm {\bf Q}}_{1 \to x}$ involve linear combinations
of $x^{r_i },i = 1,\ldots, 4$, for which $r_i = m_i$ hold.

Finally from (\ref{eq6}), (\ref{eq7}), and (\ref{eq31}) we have

\begin{equation}
\label{eq74}
f\left( x \right) = \exp \left[ {{\rm {\bf \Phi }}\left( x \right)}
\right]\,^t{\rm {\bf Q}}_{1 \to x} {\rm {\bf F}}\left( 1 \right),
\end{equation}

\noindent where ${\rm {\bf \Phi }}\left( x \right)$ has become a
constant vector, and ${\rm {\bf F}}\left( 1 \right)$ is a vector
of arbitrary constants, to be determined by boundary conditions.
However, comparing to (\ref{eq70}) shows that (\ref{eq74}) must be
indeed the general solution of (\ref{eq67}).

\subsection{Abel-Liouville-Ostogradski Formula}

Found in 1827 by Abel for second-order differential equations and by
Liouville and Ostogradsky in 1838 for the general case, the Wronskian should
satisfy [26-28]

\begin{equation}
\label{eq75}
{W}'\left( x \right) + a_{n - 1} \left( x \right)W\left( x \right) = 0,
\end{equation}

\noindent which implies that the Wronskian must be essentially a
constant if $a_{n - 1} \left( x \right) \equiv 0$. Now we show
that this equation can be readily reconstructed by DTMM.

Following the discussions in sub-section 4.4 the Wronskian
determinant takes the simple form $W = \left| {\rm {\bf D}}
\right|\,\left| {\exp \left( {x{\rm {\bf K}}} \right)}
\right|\,\,\left| {{\rm {\bf Q}}_{c \to x} } \right|\,\,\left|
{\rm {\bf F}} \right|$, where $c \in \Set$ is a constant and
\textbf{F} is the matrix of independent vectors as defined in
section 4.4. Therefore using (\ref{eq17}) and (\ref{eq39}) we find

\begin{equation}
\label{eq78}
\begin{array}{l}
 W = \prod\limits_{i > j} {\left[ {k_i \left( x \right) - k_j \left( x
\right)} \right]} \exp \left[ {x{\rm tr}\left\{ {\rm {\bf K}}
\right\}} \right]\times \\ \quad\quad \exp \left[ {ca_{n - 1}
\left( c \right) - xa_{n - 1} \left( x \right) - \int\limits_c^x
{a_{n - 1} \left( x \right)dx} } \right]\prod\limits_{i > j}
{\frac{k_i \left( c \right) - k_j \left( c \right)}{k_i \left( x
\right) - k_j \left( x \right)}} \,\left| {\rm {\bf
F}} \right| \\
 \quad = \exp \left[ {ca_{n - 1} \left( c \right)} \right]\prod\limits_{i >
j} {\left[ {k_i \left( c \right) - k_j \left( c \right)} \right]}
\,\left| {\rm {\bf F}} \right|\exp \left[ { - \int\limits_c^x
{a_{n - 1} \left( x \right)dx} } \right] \\ \quad \equiv A\exp
\left[ { - \int\limits_c^x {a_{n - 1}
\left( x \right)dx} } \right] \\
 \end{array},
\end{equation}

\noindent in which $A$ is a constant. Clearly the Wronskian given
by (\ref{eq78}) satisfies the differential equation (\ref{eq75}).

\section{Conclusions}

We have presented a new analytical formalism for solution of
linear homogeneous ordinary differential equations with variable
coefficients. The formalism is based on the definition of jump
transfer matrices and their extension into differential form. We
presented the derivative lemma and the fundamental theorem of
differential transfer matrix method, which support the exactness
of the formalism. We also discussed the linear independency of
solutions. The method is completely general, but fails in the
presence of identical roots of characteristic equation. We have
discussed a method to deal with corresponding singularities. The
main advantage of the presented method is that it deals directly
with the evolution of linear independent solutions, rather than
their derivatives. The DTMM, as described in $\S 5$, when applied
to wave propagation problems, leads to a novel approach for
understanding the behavior of physical systems.

\noindent \textbf{Acknowledgement.} The authors wish to thank
Prof. Beresford Parlett and Dr. Michael Parks at University of
California at Berkeley for valuable discussions.

\appendix

\section{Proof of (\ref{eq36})}

We prove the validity of (\ref{eq36}), by showing first that both
sides have the same derivative. From [39] the left-hand-side of
(\ref{eq36}) should obey the differential equation

\begin{equation}
\label{eqA1} \frac{d}{dx_2}\left| \exp \left[
{\int\limits_{x_1}^{x_2} {{\rm {\bf H}}\left( t \right)^{ - 1}{\rm
{\bf {H}'}}\left( t \right)dt} } \right] \right| = \left| \exp
\left[ {\int\limits_{x_1}^{x_2} {{\rm {\bf H}}\left( t \right)^{ -
1}{\rm {\bf {H}'}}\left( t \right)dt} }  \right] \right| {\rm
tr}\left\{ {\rm {\bf H}}^{ - 1}\left( x_2\right){\rm {\bf
H}}^\prime\left( x_2 \right) \right\}.
\end{equation}

\noindent Similarly, application of the derivative theorem for
determinants [32, p.178] to the right-hand-side of (\ref{eq36})
results in

\begin{equation}
\label{eqA2} \frac{d}{dx_2}\left| {\rm {\bf H}}^{ - 1}\left( x_1
\right){\rm {\bf H}}\left( x_2 \right) \right|=\frac{1}{\left|{\rm
{\bf H}}\left( x_1 \right)\right|}\frac{d\left| {\rm
{\bf H}}\left( x_2 \right) \right|}{dx_2}=\\
\frac{\left| {\rm {\bf H}}\left( x_2 \right) \right|}{\left|{\rm
{\bf H}}\left(x_1 \right)\right|}{\rm tr}\left\{{\bf
H}^{-1}(x_2){\bf H}^\prime(x_2)\right\},
\end{equation}

\noindent Therefore, both sides satisfy the same differential
equation, and thus

\begin{equation}
\label{eqA3} \left| \exp \left[ {\int\limits_{x_1}^{x_2} {{\rm
{\bf H}}\left( x \right)^{ - 1}{\rm {\bf {H}'}}\left( x \right)dx}
} \right] \right| = w(x_1) \left| {\rm {\bf H}}^{ - 1}\left( x_1
\right){\rm {\bf H}}\left( x_2 \right) \right|,
\end{equation}

\noindent where $w(\cdot)$ is a function of only $x_1$. But upon
setting $x_1=x_2$ in (\ref{eqA3}), one gets $w(\cdot)\equiv1$ and
hence the claim.

\section{Matrix Exponential of 2$\times $2 Matrices}

The difficulties in evaluation of matrix exponential have been
well known for a long time [52,53]. While computation of matrix
exponential is possible by general [54-56] and approximate [57]
methods, simple analytical expression exists only for few cases
including the Euler-Rodrigues formula [50] for $3\times 3$
skew-symmetric matrices. Here, we report an exact expansion of
matrix exponential for arbitrary $2\times 2$ square matrices.

Assume that \textbf{M} is a square matrix with arbitrary complex
elements. We define the traceless matrix $\textbf{A}=\textbf{M}-
\textstyle{1 \over 2}{\rm tr}\left\{ {\rm {\bf M}} \right\}{\rm
{\bf I}}$. Since \textbf{A} and $\textstyle{1 \over 2} {\rm
tr}\left\{ {\rm {\bf M}} \right\}{\rm {\bf I}}$ obviously commute,
one can deduce that

\begin{equation}
\label{eq76} \exp \left( {\rm {\bf M}} \right) = \exp \left(
{\textstyle{1 \over 2}{\rm tr}\left\{ {\rm {\bf M}} \right\}}
\right)\exp \left( {\rm {\bf A}} \right).
\end{equation}

\noindent Now, \textbf{A} satisfies the property ${\rm {\bf A}}^2
= \left| {\rm {\bf A}} \right|\,{\rm {\bf I}}$, so that [58]

\begin{equation}
\label{eq77} \exp \left( {\rm {\bf A}} \right) = \cos \left(
\delta \right)\;{\rm {\bf I}} + {\rm sinc}\left( \delta
\right)\;{\rm {\bf A}},
\end{equation}

\noindent in which $\delta = \sqrt { - \left| {\rm {\bf A}}
\right|}$ and $\rm{sinc}(x)=\sin(x)/x$. Finally,

\begin{equation}
\label{eq79} \exp \left( {\rm {\bf M}} \right) = \exp \left(
{\textstyle{1 \over 2}{\rm tr}\left\{ {\rm {\bf M}} \right\}}
\right)\,\left( {\cos \delta \;{\rm {\bf I}} + {\rm sinc}\delta
\;{\rm {\bf A}}} \right).
\end{equation}

\end{document}